\documentstyle[preprint,aps]{revtex}
\begin{document}
\draft
\preprint{QFUFT-6.0-12/26/94}
\title{Quantization Failure in Unified Field Theories}
\author{D.C. Galehouse}
\address{Department of Physics University of Akron, Akron, Ohio 44325}
\date{\today}
\maketitle
\begin{abstract}
Studies of geometrical theories suggest that fundmental problems of
quantization arise from the disparate usage of displacement operators.
These may be the source of a concealed inconsistency in the accepted formalism
of quantum physics.
General relativity and related theories cannot be quantized by the classical
procedure.  It is necessary to avoid
the construction of differential equations by operators applied
algebraically.  For such theories, Von Neumann's theorem concerning
hidden variables is avoided.  A specified alternative class of
gravitational-quantum-electrodynamic theories is possible.
\end{abstract}
\pacs{MS: , PACS: 3.65,4.20,4.50}
\narrowtext

\section{INTRODUCTION}
\label{sec:INTRO}

Quantum and general relativistic field theories have been studied
for many years.  The metaphysical differences between these
have not yet been resolved and a combined
theory is not available. A number of enigmas in present day
thought suggest that something is wrong.  Among these is
the persistence of wave-particle duality, the unending discussions of
measurement theory, the problem of the quantization
of gravity, and the general difficulties relating to
a principle of quantization.
An analysis by geometrical methods suggests that the basic difficulties
are caused by an incorporated internal inconsistency.
According to the rules of logic, the presence
of an internal contradiction produces arbitrary conclusions.
Even if a concealed inconsistency is not explicitly identified, it may
occur because experimental results do not agree with accepted beliefs.
To indicate the possible consequences for other theories, a short discussion
of the implications is given here. A more complete account of the geometrical
theories will be presented later.

Apparent inconsistencies often appear in
physical theories.  These occur not for failure of logic but
because of the way knowledge is obtained. Much of physics
knowledge comes from experiment.  Unfortunately, such experiments
do not directly supply a consistent set of premises.
Notwithstanding
errors in the experimental method, which are not the
subject of this article, logical consistency is often lost
because
of oversimplification, chance concurrence, or
mistaken perception.  In practice, the study of an
inconsistency or paradox is often enlightning and may reveal the
interrelationship of difficult ideas.  A more serious concern is
the existence of a real fundamental inconsistency, undiscovered,
in accepted beliefs.
These may not produce any immediate sign until analysis
or new measurement uncovers trouble in a distant or isolated
setting.  Any denial of a real inconsistency is not
productive and leads to interminable difficulties.
\footnote{A concealed inconsistency allows the proliferation of
sophistical arguments.
If the premises contain a contradiction,
a politically desirable result can always be demonstrated and the
criticism of opponents can always be disproved.  This leads to a
situation in which truth is determined by politics.}

The central argument of this paper concerns the construction of
differential equations.  To be able to calculate with
confidence, a consistent conceptual origin of the derivative is necessary.
The fundamental derivative is
defined by the process of taking limits.
\begin{equation}
{{df \over dx}=\lim_{\epsilon \to 0}{f(x+\epsilon)-f(x) \over \epsilon}}
\end{equation}
This definition implicitly describes a field in a
continuous space with a coordinate system.  Any derivative that is used for
calculation should be related to this definition.

\section{DERIVATIVES IN QUANTUM MECHANICS}
\label{sec:DQM}

In quantum mechanics, derivatives are usually
introduced by more subtle methods.
The accepted construction contains essential algebraic steps.
It is useful to follow the historical context.
Starting with Newton, motion is described by the derivative of
spatial position with respect to time.
\begin{equation}
{ F=m{d^{2}x \over dt^{2}}}
\end{equation}
This temporal equation has evolved into the more sophisticated
mathematical form that uses lagrangians and hamiltonians.
These facilitate the solution of problems containing constraints.
\footnote{Such constraints and contact forces
are now recognized as the result of
quantum phenomenology and should properly be explained by the quantum
theory of materials.}    One derives
\begin{equation}
{ {\partial \over \partial t}
\left( {\partial L \over \partial \dot x} \right) -
{\partial L \over \partial x} =0 }
\end{equation}
and
\begin{equation}
{H(p,x)=p \dot x - L \quad p={\partial L \over \partial \dot x} }
\end{equation}
where differentiations are made with respect to
to space, momentum, and other physical quantities.

The derivation of quantum equations proceeds by the process of first
quantization.  Many subtle variations and refinements exist
but the elementary scheme that follows is sufficient for the argument.
Specific suitable hamiltonians ${H=H(p,x)}$ are chosen.
These are rewritten, ${H=H(p,x)}$,
having identical appearance, but with $x$ and $p$ taken as
algebraic objects rather than as fields.
Into this expression, which must be carefully arranged,
derivative operators are inserted in place of the momentum and energy
symbols.  These derivatives
${p \to { \hbar \over i}{\partial \over \partial x }}$ and
${E \to i \hbar {\partial \over \partial t} }$
are introduced algebraically
without consideration for the elementary definition~\cite{DD}.
They must not appear in difficult places and they must be normal ordered
by a tenuous set of rules.  This
assembled operator can then be applied to an hypothesized wave function.
\begin{equation}
{i \hbar {\partial \over \partial t}\psi
=H\left( {\hbar \over i}{\partial \over \partial x},x\right) \psi}
\end{equation}
The resulting equations agree well with experiment, and in an historical
context,demonstrate unequivocal success.

\section{DERIVATIVES IN GENERAL RELATIVITY}
\label{sec:DGR}

In contrast, derivatives in general relativity are developed from
a sense of displacement that is equivalent to the elementary
definition~\cite{ANDERSON}.
To obtain mathematical consistency in
curvilinear coordinate systems, a coefficient of connection
${\Gamma^{\nu}_{\beta \mu}}$ must be defined and used.
\begin{equation}
V^{\nu}( x^{\mu}+\Delta x^{\mu})  =
V^{\nu}( x^{\mu})+{\partial V^{\nu} \over \partial x^{\mu}}\Delta x^{\mu}+
\Gamma^{\nu}_{\beta \mu} V^{\beta}\Delta x^{\mu}.
\end{equation}
The connection ${\Gamma^{\nu}_{\beta \mu}}$  is set equal to the
christoffel symbol in general relativity
and in other Riemannian geometries.
\begin{equation}
\Gamma^{\nu}_{\beta\mu}={1 \over 2}g^{\nu \lambda}
 \left( {\partial g_{\mu \lambda} \over \partial x^{\beta}}+
 {\partial g_{\lambda \beta} \over \partial x^{\mu}}-
 {\partial g_{\beta \mu} \over \partial x^{\lambda}}\right)
\end{equation}

This construction
is essential because of the use of geodesics as a fundamental
description of gravitational motion.
To define a coefficient of connection, common derivatives
of a metric or other tensor
are essential and unavoidable.
For non-Riemannian geometries, an additional tensor is added to the
christoffel symbol in forming the connection.

\section{COMBINATIONS OF QUANTUM AND RELATIVISTIC DERIVATIVES}
\label{sec:QAR}

The quantum and relativistic methods are antagonistic.
A classical theory of mechanics, suitable
for the application of first quantization, must have a momentum vector
${p^{\mu}}$ for even one
simple particle.  This classical momentum, a first order relativistic
vector, must transform correctly under arbitrary coordinate transformations.
Consequently, it has a displacement which must be defined with the aid of a
coefficient of connection.

Upon application of the first quantization procedure, a physical
momentum is replaced by a momentum
operator.  That operator has already been used in the definition of the
classical covariance of the quantity it repaces.  Moreover,
the derivatives associated with the connection have undefined
properties during the process of constructing an algebraic hamiltonian.
Those contained within the definition of the coefficient
of connection cannot be treated as algebraic
quantities.   Following this, the selection of a completed operator
for the wave function seems problematical.
The introduction of derivatives in different ways, one
from the quantum operator, one from the coefficient of connection,
does not establish their equivalence.
Nor is it reasonable to suppose that the result
can be made covariant by any proper mathematical manipulation.    The
absence of a unified concept of physical derivative remains an impasse
to progress.  Many attempts to combine quantum mechanics and
general relativity seem to fail because of this  inconsistency in
their respective mathematical foundations.

\section{IS A UNIFIED THEORY POSSIBLE}
\label{sec:UFT?}

The coexistence of separate incompatible concepts of differentiation
has been accepted since the 1920's.  The mathematical problem has
not been resolved nor fully discussed.

In principle it may be possible to avoid these problems by supposing
an exclusively algebraic structure without
any initial geometrical presumptions.
Geodesics, tensors, curvatures and
all but the simplest sorts of coordinates would be avoided.
Then, after algebraic
first quantization, the phenomenology of general relativity would have to be
derived in some approximation by appropriate manipulations.
Such a method may be possible, but is not yet a convincing reality.

The alternative approach suggested here is to construct quantum
theories without using first quantization.
There is no presupposed classical theory.
One studies suitable
geometrical invariants to find such expressions as might
describe quantum mechanics with or without
other fields and interactions.  In principle,
such a theory may be derivable from a lagrangian; but, without the
availability of a classical momentum, the standard rules for defining
such a lagrangian are inadequate.   The usual methods for defining models
fail and an entirely different approach is required.

\section{A GEOMETRICAL THEORY OF QUANTUM MECHANICS}
\label{sec:GTQM}

As an example of such a direct quantum construction,
a theory that is consistent with
these conditions can be constructed from general relativity.
Using the minimal substitution as a transformation of derivatives,
${{\partial \over \partial x^{\mu}}  \to
{\partial \over \partial x^{\mu} }-
{ie \over \hbar c}A_{\mu}}$
and applying it naively to the usual Riemannian connections
of general relativity, there results
\begin{equation}
\Gamma^{\prime\nu}_{\mu \beta}  =
{\nu \brace \mu\beta}-
{ie \over 2 \hbar c}\left(
\delta^{\nu}_{\mu}A_{\beta}+
\delta^{\nu}_{\beta}A_{\nu}-
g_{\mu \beta}g^{\nu \lambda}A_{\lambda}\right) .
\end{equation}
There is no apriori reason to believe that this will work an generate
as useful theory.  However,
if the change is applied uniformly, then at least
the continued consistency of the infinitesimal displacements is assured.
The result is non-Riemannian but precisely of the form supposed by
Weyl~\cite{WEYL} in his early unified field theory.
The numerical factor is suitable for quantum
scale events.  The known properties of Weyl's theory guarantee
covariance.   A more complete investigation
shows that a fairly complete quantum structure is contained within
the geometry.  Quantum phenomena are predicted without quantization.

\section{BETTER UNIFIED FIELD THEORIES}
\label{sec:BUFT}

More complicated but also more complete field theories are possible.
One can at least hope to
include field and motion equations for quantum mechanics,
general relativity and electrodynamics.  Such a system should
have geodesic trajectories and consistent displacement operators.

The known fundamental geometrical theories incorporate constructions
that cause special problems.  The derivatives in the connections are
equivalent to momentum operators.
A further direct application of quantization to these theories
is likely to lead to over quantization.  The degree of
the differential equations will be too high leading at least to
ghost solutions and extraneous interdependency of physical
effects~\cite{OQFT}.
A consistent interpretation then becomes difficult.

A terrifying set of conclusions is suggested. All classical relativistic
theories that incorporate curvilinear coordinates must be incorrect because
they contain suppressed quantum fields.  None of them can be properly
quantized by any conventional procedure.  Quantum general relativistic
theories cannot be constructed from classical physics but require a separate
self-consistent set of premises.  IT is suggested that the
hamiltonian approach proposed by Dirac~\cite{DIRAC} and pursued by
others~\cite{HQ} will continue to incur difficulties.

The internal construction of general relativity suggests that any other
general covariant theory of interaction will likely be of
comparable complexity.  These theories must contain connections
and derivatives, implicit or explicit.  Such a relativistic quantum
theory must imply connections and consequently
cannot have a classical precursor.  The general accepted methods to search
for covariant quantum field theories fail systematically. A particular
construction my
succeed fortuitously, but a better method is to use consistent geometrical
derivations from the beginning.

\section{INAPPLICABILITY OF VON NEUMANN'S THEOREM}
\label{sec:IVNT}

It is important to note that
Von Neumann's theorem concerning hidden variables does not  apply to
fully geometrical theories of this type~\cite{SUSQM}.   Because
the momentum operator is not a physical quantity but
a plain mathematical symbol,
the epistemological association with classical physics
is dropped and observations are based on geodesics or other geometrical
quantities.  There is no conventional
measurement theory.   Such a
theory, if it were to have hidden variables, could only contain such
quantities as transform properly. Bare
differential operators, ${p_{(op)}}$, do not qualify.
More generally, any theory that
does not associate physical quantities with a momentum operator
or that does not attach hidden variables to differential quantum
observables, abrogates the conditions of
Von Neumann's theorem.
The physical identification of curvilinear coordinates with particle
motion avoids measurement theory.  The requirement of mathematical
consistency for the derivatives bypasses conventional quantization.

\section{SUMMARY AND CONCLUSION}
\label{sec:CONC}

The historical failure to combine quantum mechanics
with relativistic theories is argued to be due to fundamental
failure in the meta-mathematical structure of derivatives.
While the synthesis of gravitation, quantum mechanics,
and electrodynamics may be possible,
the equations should be written in terms of geometrical invariants.
They must use displacement operators self-consistently.
Measurement theory is avoided and Von Neumann's theorem does not
apply. Such geometrical theories
cannot be constructed quantum free nor can they be derived
by first quantization.

It may also be possible to have a geometry of interaction.
Noting the work of Kaluza~\cite{KALUZA}, five dimensional theories may
have the  potential to incorporate geometrically  based inhomogeneous field
equations. Various authors have attributed either the field or
motion equations to five dimensional effects~\cite{FDT}.
Although these are not widely
accepted, some of the difficulties may be because of the irreconcilable
differences between latent quantum effects and classical interpretations.
The interaction constants may be contained in the geometry.  Those
describing the gravitational and electromagnetic field should have
geometrical interpretations. Unfortunately,  all workable theories
of this type challenge the formalism and interpretation of conventional
quantum mechanics.  The investigation of these remarkable
mathematical structures may lead to predictions that otherwise cannot
be formulated.

\acknowledgements
I would like to thank Dr. E. von Meerwall for translating an occasional German
article and B. Galehouse for helping with the typesetting.

\end{document}